\documentclass[pra,a4paper,preprint,tightenlines,showpacs]{revtex4-1}
\usepackage[utf8]{inputenc}
\usepackage{amsmath,graphicx,hyperref}
\usepackage{braket}
\usepackage[italic]{hepnames}

\makeatletter
\renewcommand{\@caption@fignum@sep}{ (Color online). }
\makeatother

\begin{document}
\title{Neutral kaons as an open quantum system in a second
  quantization approach}

\author{Kordian Andrzej Smoli\'nski}

\email{K.A.Smolinski@merlin.phys.uni.lodz.pl}

\affiliation{Department of Theoretical Physics, 
  Faculty of Physics and Applied Computer Sciences, 
  University of Lodz, ul.\ Pomorska
  149/153, 90-236 \L\'od\'z, Poland}

\date{September 24, 2015}

\begin{abstract}
  We have shown that it is possible to formulate the consistent and
  probability-preserving description of the $CP$-symmetry-violating
  evolution of a system of decaying particles.  This has been done
  within the framework of quantum mechanics of open systems.  This
  approach allows the description of both the exponential decay and
  flavour oscillations.  We have solved explicitly the
  Kossakowski-Lindblad master equation for a system of particles with
  violated $CP$ symmetry and found the evolution of \emph{any}
  observable bilinear in creation and annihilation operators.  The
  choice of a concrete observable can be done by the proper choice of
  initial conditions for the system of differential equations.  We
  have calculated the evolution as well as mean values of the
  observables most interesting from the physical point of view, and we
  have found their lowest order difference with the $CP$-preserved
  values.
\end{abstract}


\maketitle

\section{Introduction}
\label{sec:intro}

In recent years Bell inequalities \cite{bell64,*chsh,*clauser78} were
tested in systems of correlated neutral $\PK$
\cite{uchiyama97,*bramon99,*foadi99a,*dalitz01,
  *bertlmann01,*bertlmann01a,*bramon02,*genovese04,bertlmann03} or
$\PB$ mesons \cite{go04,*go07:abb,*bramon05a}.  However, the quantum
mechanical analysis of such systems encounters an important
difficulty, namely, the irreversibility of time-evolution of unstable
particles.  Only the complete system consisting of the decaying
particle as well as the decay products undergoes unitary evolution,
which is actually described by quantum field theory.  But in
Einstein-Podolsky-Rosen (EPR) correlation experiments
\cite{einstein35,*bohm51}, it is more useful to consider the decaying
particles only, neglecting the evolution of decay products.  In the
usual approach one introduces a non-Hermitian Hamiltonian, as it was
done in the classical works of Weisskopf and Wigner
\cite{weisskopf30a,*weisskopf30b}.  However, the non-Hermitian
Hamiltonian does not provid an unambiguous way of calculating the
probability of finding the system consisting of a few such particles
in a given state after the measurement (such an approach causes also
other ambiguities, see e.g.\ \cite{durt13} for a discussion).  Indeed,
using this approach, we must reinterpret quantum mechanical results
with the help of probability theory, especially if we would like to
analyze correlation experiments with unstable particles, namely,
neutral $\PK$ or $\PB$ mesons \cite{ellis96,*bernabeu13}.  From the
quantum mechanical point of view, this means that unit trace and
positivity of the density matrix for the system under consideration
must be preserved
\cite{benatti96,*benatti97a,*benatti97b,*benatti98,*benatti98a,
  *benatti98b:full,*benatti01}.  Fortunately, it is possible to
resolve the above-mentioned problems using the open quantum systems
theory \cite{alicki01,*breuer02}.  The idea that unstable particles
can be treated as open quantum systems was proposed first by R.~Alicki
\cite{alicki77,*alicki78,*alicki07:full} and was developed by various
authors in different contexts (see e.g.\ \cite{weron85,liu04}).
Recently, this approach was successfully applied to the systems of
particles with flavour oscillations (like in the case of neutral kaons
or $\PB$ mesons) \cite{bertlmann06,caban05a,smolinski14}, and it has
been used successfully in the description of EPR correlations and
evolution of entanglement in $\PKzero\APKzero$ system
\cite{caban06a:abb,caban07}.

In this theory the evolution of a quantum system is described by the
master equation \cite{kossakowski72,*gorini76,*lindblad76}, which can
be treated as the replacement for either von Neumann or Heisenberg
equations, for the picture of quantum mechanics in use.  Here, we
follow the approach presented in \cite{smolinski14}, where systems
with an arbitrary number of particles were described with the use of
the second quantization formalism, which seems to be the most natural
language for a system with varying number of particles.

The paper is organized as follows.  In Sect.~\ref{sec:states} we
introduce the notation and conventions used through the paper.  In
Sect.~\ref{sec:master} we introduce the master equation for a system
of neutral kaons with violated $CP$ symmetry and we find the evolution
of the observables in the Heisenberg picture, and, in
Sect.~\ref{sec:evolution}, we analyze the evolution of the total
number of particles and strangeness observable as well as the numbers
of each of neutral kaon flavor.  Finally, we conclude the paper in
Sec.~\ref{sec:concl}.

\section{Preliminaria}
\label{sec:states}

We start with introducing some notations, conventions and definitions
used throughout the paper, as well as give experimental values of some
parameters important in the description of the systems of $\PKzero$
and $\APKzero$.

We define one-particle states $\ket{\PKzero}$ and $\ket{\APKzero}$ as
\begin{subequations}
  \label{eq:K00}
  \begin{align}
    \ket{\PKzero} &= a^\dag \ket{0}\,, \\
    \ket{\APKzero} &= b^\dag \ket{0}\,,
  \end{align}
\end{subequations}
where $a$ and $b$ fulfills the usual canonical commutation relations
\begin{subequations}
  \begin{gather}
    [a, a^\dag] = [b, b^\dag] = 1\,, \\
    [a, b] = [a^\dag, b^\dag] = 0\,.
  \end{gather}
\end{subequations}
States $\ket{\PKzero}$ and $\ket{\APKzero}$ are orthonormal, i.e.\
\begin{subequations}
  \begin{gather}
    \braket{\PKzero|\PKzero} = \braket{\APKzero|\APKzero} = 1\,,\\
    \braket{\PKzero|\APKzero} = 0\,,
  \end{gather}
\end{subequations}
and they are eigenstates of the strangeness operator $S = a^\dag a -
b^\dag b$:
\begin{subequations}
  \begin{align}
    S \ket{\PKzero} &= \ket{\PKzero}\,, \\
    S \ket{\APKzero} &= -\ket{\APKzero}\,.
  \end{align}
\end{subequations}

For our convenience we will use the following short-hand notation for
multiparticle states:
\begin{equation}
  \ket{\#\PKzero=n,\#\APKzero=\bar{n}} \equiv \ket{n,\bar{n}}\,.
\end{equation}
Therefore
\begin{equation}
  \label{eq:nn'}
  \ket{n,\bar{n}} = \frac{1}{\sqrt{n! \bar{n}!}} \left(a^\dag\right)^n 
  \left(b^\dag\right)^{\bar{n}} \ket{0}
\end{equation}
and
\begin{equation}
  \label{eq:Snn'}
  S \ket{n, \bar{n}} = (n - \bar{n}) \ket{n, \bar{n}}\,.
\end{equation}


However, the states $\ket{\PKzero}$ and $\ket{\APKzero}$ are not
eigenstates of time evolution.  Actually, under time evolution the
``well behaving'' states are $\ket{\PKshort}$ and $\ket{\PKlong}$,
defined as follows
\begin{subequations}
  \label{eq:KSL}
  \begin{align}
    \ket{\PKshort} &= p \ket{\PKzero} + q \ket{\APKzero}\,, \\
    \ket{\PKlong} &= p \ket{\PKzero} - q \ket{\APKzero}\,,
  \end{align}
\end{subequations}
with mean life times $\tau_S = 0.8954 \times 10^{-10}\,\mathrm{s}$ and
$\tau_L = 5.116 \times 10^{-8}\,\mathrm{s}$, respectively
\cite{olive14}.  Moreover, if $p \neq q$ then $CP$ symmetry is
violated.  Indeed,
\begin{equation}
  \label{eq:pq}
  \frac{p}{q} = \frac{1 + \tilde\epsilon}{1 - \tilde\epsilon}
\end{equation}
 with
\begin{subequations}
  \label{eq:AL}
  \begin{align}
    |p|^2 + |q|^2 &= 1\,, \\
    |p|^2 - |q|^2 &= A_L\,,
  \end{align}
\end{subequations}
and $A_L = 2 \operatorname{Re}(\epsilon)/(1 + |\epsilon|^2)$ is a
measure of violation of $CP$ symmetry (hereafter we use the so-called
Wu-Yang phase convention \cite{wu64} in which $\tilde\epsilon =
\epsilon$).  Experimentally obtained values are $A_L = 0.332\%$ and
$|\epsilon| = 2.228 \times 10^{-3}$ \cite{olive14}.  Notice also that
the basis~\eqref{eq:KSL} is not orthogonal since
\begin{equation}
  \label{eq:braketSL}
  \braket{\PKshort|\PKlong} = A_L\,.
\end{equation}

\section{Master equation}
\label{sec:master}

The aim of this section is to find the evolution of the multiparticle
system of $\PKzero$ and $\APKzero$.  We know that the evolution of a
single neutral kaon can be described as an open system obeying the
Kossakowski-Lindblad master equation \cite{caban05a}.  In this case
the one-particle Hamiltonian for this master equation was written in
the base~\eqref{eq:KSL} and reads
\begin{subequations}
  \label{eq:me1SL}
  \begin{multline}
    \label{eq:H1SL}
    H = \frac{1}{1 - A_L^2} \left\{m_S \ket{\PKshort}\bra{\PKshort} +
      m_L \ket{\PKlong}\bra{\PKlong}\vphantom{\tfrac{i}{4}
        \Delta\Gamma}\right.\\
    \left.- A_L \left[\left(m - \tfrac{i}{4} \Delta\Gamma\right)
        \ket{\PKshort}\bra{\PKlong} + \left(m + \tfrac{i}{4}
          \Delta\Gamma\right)
        \ket{\PKlong}\bra{\PKshort}\right]\right\}\,,
  \end{multline}
  while Lindbladians are of the form
  \begin{align}
    L_1 &= \frac{1}{1 - A_L^2} \sqrt{\Gamma_S - A_L^2 \frac{\Gamma^2 +
        \Delta{m}^2}{\Gamma_L}} \left(\ket{0}\bra{\PKshort} - A_L
      \ket{0}\bra{\PKlong}\right)\,,\\
    L_2 &= \frac{1}{1 - A_L^2} \left[\left(\sqrt{\Gamma_L} - A_L^2
        \frac{\Gamma - i \Delta{m}}{\sqrt{\Gamma_L}}\right)
      \ket{0}\bra{\PKlong} 
        - A_L \left(\sqrt{\Gamma_L} - \frac{\Gamma
          - i \Delta{m}}{\sqrt{\Gamma_L}}\right)
      \ket{0}\bra{\PKshort}\right]\,,
  \end{align}
\end{subequations}
and $K \equiv -(L_1^\dag L_1 + L_2^\dag L_2)/2$, where $m = (m_S +
m_L)/2 = 497.614\, \mathrm{MeV}/c^2$ is the $\PKzero$ mean mass,
$\Delta{m} = m_L - m_S = 0.5293 \times 10^{10}\, \hbar/\mathrm{s}$
\cite{olive14}, $\Gamma = (\Gamma_S + \Gamma_L)/2$, and $\Delta\Gamma
= \Gamma_S - \Gamma_L$.

The state of a single particle system under the evolution given by the
master equation with operators~\eqref{eq:me1SL} is in general a mixed
state of a single particle and vacuum (see \cite{caban05a}).  After
the projection of this state on the single-particle sector one gets
the state obtained by the non-trace-preserving equations, considered
usually in the literature (see e.g.\ \cite{bertlmann03}).

\subsection{Master equation in second quantization formalism}
\label{sec:master2q}

Now, let us go to the description of a multiparticle system.  We
achieve this by the replacement of intertwining operators with the
appropriate combinations of annihilation and creation operators, as
was done in~\cite{smolinski14}.  In the base~\eqref{eq:K00} this
Hamiltonian and these Lindbladians take the form
\begin{subequations}
  \label{eq:me00}
  \begin{equation}
    \label{eq:H100}
    H = m \left(a^\dag a + b^\dag b\right) - \frac{p q^*}{1 - A_L^2}
    \left(\Delta{m} + \tfrac{i}{2} A_L
      \Delta\Gamma\right) a^\dag b 
    - \frac{q p^*}{1 - A_L^2} \left(\Delta{m} - \tfrac{i}{2} A_L
      \Delta\Gamma\right) b^\dag a
  \end{equation}
  and
  \begin{align}
    L_1 &= \sqrt{\Gamma_S - A_L^2 \frac{\Gamma^2 +
        \Delta{m}^2}{\Gamma_L}} \left(\frac{p^*}{1+A_L} a +
      \frac{q^*}{1-A_L} b\right)\,,\\
    L_2 &= \frac{p^*}{1+A_L} \left(\sqrt{\Gamma_L} + A_L \frac{\Gamma
        - i \Delta{m}}{\sqrt{\Gamma_L}}\right) a 
    - \frac{q^*}{1-A_L} \left(\sqrt{\Gamma_L} - A_L
      \frac{\Gamma - i \Delta{m}}{\sqrt{\Gamma_L}}\right) b\,,\\
    K &= -\tfrac{1}{2} \Gamma \left(a^\dag a + b^\dag b\right) -
    \frac{p q^*}{1 - A_L^2} \left(\tfrac{1}{2}\Delta\Gamma - i A_L
      \Delta{m}\right) a^\dag b 
    - \frac{q p^*}{1 - A_L^2} \left(\tfrac{1}{2}\Delta\Gamma + i A_L
      \Delta{m}\right) b^\dag a
  \end{align}
\end{subequations}

Let us consider a single particle state described by a density matrix
of the form
\begin{equation}
  \label{eq:rho}
  \rho = p_1 \ket{\PKzero} \bra{\PKzero} + p_2 \ket{\APKzero}
  \bra{\APKzero} + w \ket{\PKzero} \bra{\APKzero} + w^* \ket{\APKzero}
  \bra{\PKzero} + (1 - p_1 - p_2) \ket{0} \bra{0}
\end{equation}
with $0 \leq p_1 \leq 1$, $0 \leq p_2 \leq 1$, $0 \leq p_1 + p_2 \leq
1$, and $|w|^2 \leq p_1 p_2$.  The evolution of this state in the
Schrödinger picture is governed by the master equation
\begin{equation}
  \label{eq:meSch}
  \partial_t \rho(t) = -i [H, \rho(t)] + \{K, \rho(t)\} 
  + \sum_i L_i \rho(t) L_i^\dag\,.
\end{equation}
By a straightforward calculation, using relation~\eqref{eq:KSL}
between the bases $\{\ket{\PKshort}, \ket{\PKlong}\}$ and
$\{\ket{\PKzero}, \ket{\APKzero}\}$, one can check that the choices~of
Eqs.~\eqref{eq:me1SL} and~\eqref{eq:me00} for operators appearing in
Eq.~\eqref{eq:meSch} lead to the same evolution equation for
$\rho(t)$.

Moreover, the evolution of the two particle system obtained from
Eqs.~\eqref{eq:me00} and~\eqref{eq:meSch} lead to the evolution of the
two particle state obtained by taking a symmetrized (the particles are
indistinguishable) tensor product of single particle evolution (see
\cite{caban06a:abb}).

Now, consider the Kossakowski-Lindblad master equation for the
evolution of observable $\Omega(t)$ in the Heisenberg picture
\begin{equation}
  \partial_t\Omega(t) = \mathcal{L}[\Omega(t)]\,,
\end{equation}
where
\begin{equation}
  \label{eq:LOmega}
  \mathcal{L}[\Omega(t)] = i [H, \Omega(t)] 
  + \frac{1}{2} \sum_i \left\{[L_i^\dag, \Omega(t)] L_i 
    + L_i^\dag [\Omega(t), L_i]\right\}\,,
\end{equation}
and we choose $H$ and $L_i$ ($i = 1, 2$) in the form~\eqref{eq:me00}.
We assume that $\Omega(t)$ can be written as a bilinear form in
annihilation and creation operators
\begin{equation}
  \label{eq:Omega}
  \Omega(t) = \omega_{aa}(t) a^\dag a + \omega_{ab}(t) a^\dag b 
  + \omega_{ba}(t) b^\dag a + \omega_{bb}(t) b^\dag b\,,
\end{equation}
with the condition $\omega_{ba}(t) = \omega_{ab}^*(t)$ to guarantee
that $\Omega(t)$ is Hermitian.

\subsection{General solution}
\label{sec:general}

Here we want to find the evolution of any observable $\Omega(t)$ which
is bilinear in annihilation and creation operators.  To achieve this
aim, we begin with the following observation:
\begin{subequations}
  \label{eq:Laa+}
  \begin{align}
    \mathcal{L}[a^\dag a] &= -\Gamma a^\dag a - \frac{p q^*}{1 - A_L}
    \left(\tfrac{1}{2} \Delta\Gamma - i \Delta{m}\right) a^\dag b
    - \frac{q p^*}{1 - A_L} \left(\tfrac{1}{2} \Delta\Gamma + i
      \Delta{m}\right) b^\dag a\,, \\
    \mathcal{L}[a^\dag b] &= -\Gamma a^\dag b - \frac{q p^*}{1 + A_L}
    \left(\tfrac{1}{2} \Delta\Gamma - i \Delta{m}\right) a^\dag a
    - \frac{q p^*}{1 - A_L} \left(\tfrac{1}{2} \Delta\Gamma + i
      \Delta{m}\right) b^\dag b\,, \\
    \mathcal{L}[b^\dag a] &= -\Gamma b^\dag a - \frac{p q^*}{1 + A_L}
    \left(\tfrac{1}{2} \Delta\Gamma + i \Delta{m}\right) a^\dag a
    - \frac{p q^*}{1 - A_L} \left(\tfrac{1}{2} \Delta\Gamma - i
      \Delta{m}\right) b^\dag b\,, \\
    \mathcal{L}[b^\dag b] &= -\Gamma b^\dag b - \frac{p q^*}{1 + A_L}
    \left(\tfrac{1}{2} \Delta\Gamma + i \Delta{m}\right) a^\dag b
    - \frac{q p^*}{1 + A_L} \left(\tfrac{1}{2} \Delta\Gamma - i
      \Delta{m}\right) b^\dag a\,.
  \end{align}
\end{subequations}
We see that if $\Omega(t)$ is of the form~\eqref{eq:Omega} at a given
moment of time $t_0$, it must preserve this form for all the time $t
\geq t_0$.

If we take into account the linearity of $\Omega(t)$ and linear
independence of operators $a^\dag a$, $a^\dag b$, $b^\dag a$, and
$b^\dag b$, we get the following system of first-order differential
equations for $\omega$:
\begin{subequations}
  \label{eq:domega}
  \begin{align}
    \dot\omega_{aa}(t) &= -\Gamma \omega_{aa}(t) - \frac{q p^*}{1 +
      A_L} \left(\tfrac{1}{2} \Delta\Gamma - i \Delta{m}\right)
    \omega_{ab}(t) 
    - \frac{p q^*}{1 + A_L} \left(\tfrac{1}{2}
      \Delta\Gamma + i \Delta{m}\right) \omega_{ba}(t)\,,\\
    \dot\omega_{ab}(t) &= -\frac{p q^*}{1 - A_L} \left(\tfrac{1}{2}
      \Delta\Gamma - i \Delta{m}\right) \omega_{aa}(t) - \Gamma
    \omega_{ab}(t) 
    - \frac{p q^*}{1 + A_L} \left(\tfrac{1}{2}
      \Delta\Gamma + i \Delta{m}\right) \omega_{bb}(t)\,,\\
    \dot\omega_{ba}(t) &= -\frac{q p^*}{1 - A_L} \left(\tfrac{1}{2}
      \Delta\Gamma + i \Delta{m}\right) \omega_{aa}(t) - \Gamma
    \omega_{ba}(t) 
    - \frac{p q^*}{1 - A_L} \left(\tfrac{1}{2}
      \Delta\Gamma - i \Delta{m}\right) \omega_{bb}(t)\,,\\
    \dot\omega_{bb}(t) &= -\frac{q p^*}{1 - A_L} \left(\tfrac{1}{2}
      \Delta\Gamma + i \Delta{m}\right) \omega_{ab}(t) 
    - \frac{p q^*}{1 - A_L} \left(\tfrac{1}{2} \Delta\Gamma - i
      \Delta{m}\right) \omega_{ba}(t) - \Gamma \omega_{bb}(t)\,.
  \end{align}
\end{subequations}

Using straightforward methods we can find that the general solution
of~\eqref{eq:domega} is of the form
\begin{widetext}
\begin{subequations}
  \label{eq:omega}
  \begin{multline}
    \omega_{aa}(t) = \frac{e^{-\Gamma t}}{2}
    \left\{\left[\cosh\left(\tfrac{1}{2}\Delta\Gamma t\right) +
        \cos(\Delta{m} t)\right] \omega_{aa}(0) \right. \\   \left.
      - \frac{q}{p} \left[\sinh\left(\tfrac{1}{2}\Delta\Gamma
          t\right) - i
        \sin(\Delta{m} t)\right] \omega_{ab}(0) 
      - \frac{1 - A_L}{1 + A_L} \frac{p}{q}
      \left[\sinh\left(\tfrac{1}{2}\Delta\Gamma t\right) + i
        \sin(\Delta{m} t)\right] \omega_{ba}(0) \right.\\    \left. 
      + \frac{1 - A_L}{1 + A_L}
      \left[\cosh\left(\tfrac{1}{2}\Delta\Gamma t\right) -
        \cos(\Delta{m} t)\right] \omega_{bb}(0)\right\}\,,
  \end{multline}
  \begin{multline}
    \omega_{ab}(t) = \frac{e^{-\Gamma t}}{2} \left\{-\frac{p}{q}
      \left[\sinh\left(\tfrac{1}{2}\Delta\Gamma t\right) - i
        \sin(\Delta{m} t)\right] \omega_{aa}(0) \right. \\ \left.
      + \left[\cosh\left(\tfrac{1}{2}\Delta\Gamma t\right) +
        \cos(\Delta{m} t)\right] \omega_{ab}(0) 
      + \frac{1 - A_L}{1 + A_L} \left(\frac{p}{q}\right)^2
      \left[\cosh\left(\tfrac{1}{2}\Delta\Gamma t\right) -
        \cos(\Delta{m} t)\right] \omega_{ba}(0) \right. \\ \left.
      - \frac{1 - A_L}{1 + A_L} \frac{p}{q}
      \left[\sinh\left(\tfrac{1}{2}\Delta\Gamma t\right) + i
        \sin(\Delta{m} t)\right] \omega_{bb}(0)\right\}\,,
  \end{multline}
  \begin{multline}
    \omega_{ba}(t) = \frac{e^{-\Gamma t}}{2} \left\{-\frac{1 + A_L}{1
        - A_L} \frac{q}{p} \left[\sinh\left(\tfrac{1}{2}\Delta\Gamma
          t\right) + i \sin(\Delta{m} t)\right] \omega_{aa}(0) \right.\\
    \left.  + \frac{1 + A_L}{1 - A_L} \left(\frac{q}{p}\right)^2
      \left[\cosh\left(\tfrac{1}{2}\Delta\Gamma t\right) -
        \cos(\Delta{m} t)\right] \omega_{ab}(0) 
      + \left[\cosh\left(\tfrac{1}{2}\Delta\Gamma t\right) +
        \cos(\Delta{m} t)\right] \omega_{ba}(0) \right.\\ \left.  -
      \frac{q}{p} \left[\sinh\left(\tfrac{1}{2}\Delta\Gamma t\right) -
        i \sin(\Delta{m} t)\right] \omega_{bb}(0)\right\}\,,
  \end{multline}
  \begin{multline}
    \omega_{bb}(t) = \frac{e^{-\Gamma t}}{2} \left\{\frac{1 + A_L}{1 -
        A_L} \left[\cosh\left(\tfrac{1}{2}\Delta\Gamma t\right) -
        \cos(\Delta{m} t)\right] \omega_{aa}(0) \right. \\ \left.  -
      \frac{1 + A_L}{1 - A_L} \frac{q}{p}
      \left[\sinh\left(\tfrac{1}{2}\Delta\Gamma t\right) + i
        \sin(\Delta{m} t)\right] \omega_{ab}(0) 
      - \frac{p}{q} \left[\sinh\left(\tfrac{1}{2}\Delta\Gamma t\right)
        - i \sin(\Delta{m} t)\right] \omega_{ba}(0) \right. \\ \left.
      + \left[\cosh\left(\tfrac{1}{2}\Delta\Gamma t\right) +
        \cos(\Delta{m} t)\right] \omega_{bb}(0)\right\}\,.
  \end{multline}
\end{subequations}
\end{widetext}
The Hermicity of $\Omega(t)$ is preserved (i.e.\ the condition
$\omega_{ab}(0) = \omega_{ba}^*(0)$ implies $\omega_{ab}(t) =
\omega_{ba}^*(t)$ for all the time $t \geq 0$), since from
Eqs.~\eqref{eq:pq} and~\eqref{eq:AL} we have
\begin{equation}
  \left(\frac{p}{q}\right)^* = \frac{1 + A_L}{1 - A_L} 
  \frac{q}{p}\,.
\end{equation}

Now let us observe that the different choices of initial conditions
for $\omega$'s allow us to get the time evolution of different
physically interesting observables of the form~\eqref{eq:Omega}.

\section{Evolution of the observables and their averages }
\label{sec:evolution}

Now we are prepared to find the evolution of some physically
interesting observables, as well as the time dependence of their mean
values.  We begin with the total number of particles and strangeness
observables, next we analyze the numbers of each neutral kaon flavor.

\subsection{Total number of particles and strangeness}
\label{sec:NS}

\begin{figure}
  \centering
  \includegraphics[width=\columnwidth]{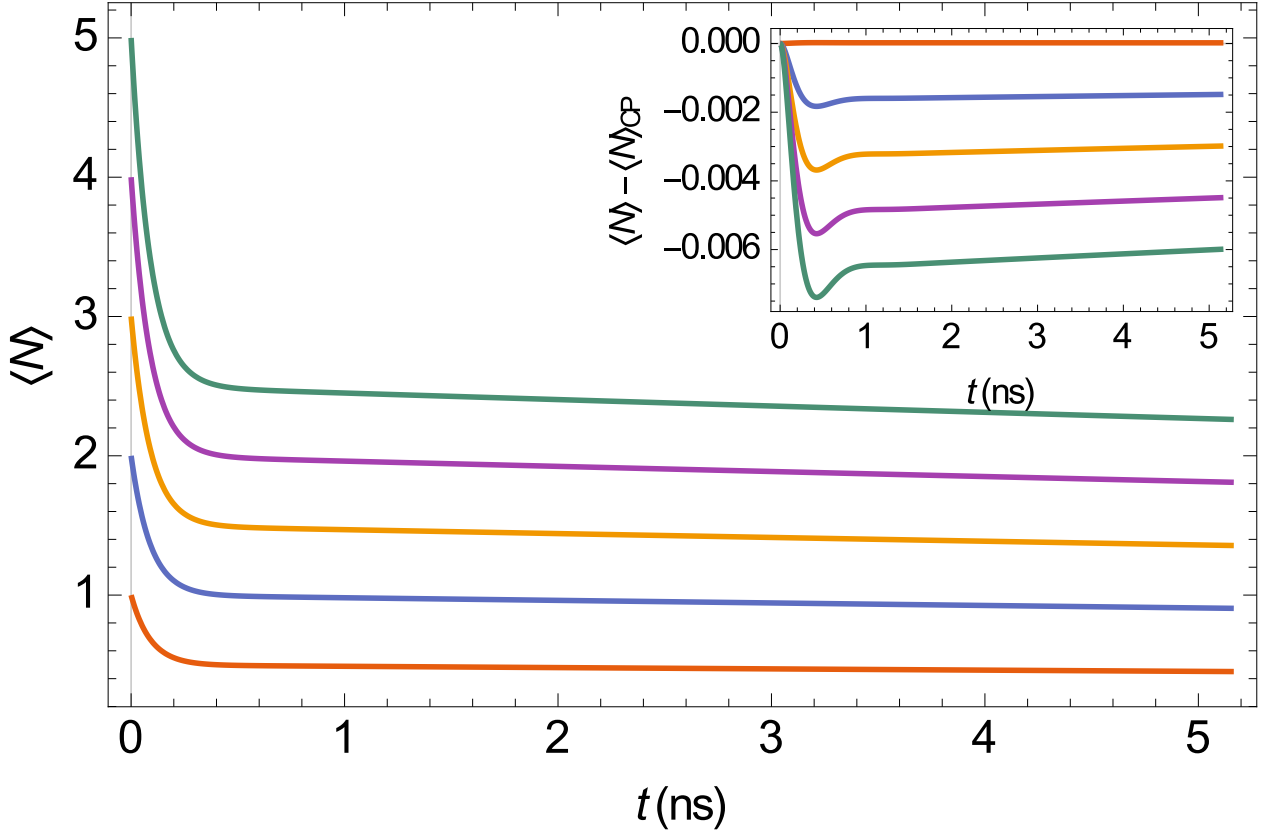}
  \caption{Mean number of particles vs time for $n + \bar{n} =
    1,\ldots, 5$ (bottom-up) and the difference between
    $\left<N(t)\right>$ for $CP$-violated and $CP$-preserved case vs
    time for $n + \bar{n} = 4$ and $n - \bar{n} = 0,\ldots, 4$
    (top-down).}
  \label{fig:N}
\end{figure}

For the total number of particles observable $N$, we have $N(0) =
a^\dagger a + b^\dagger b$, so $\omega_{aa}(0) = \omega_{bb}(0) = 1$
and $\omega_{ab}(0) = \omega_{ba}(0) = 0$, and finally we get
\begin{multline}
  N(t) = e^{-\Gamma t} \left\{\frac{1}{1 + A_L}
    \left[\cosh\left(\tfrac{1}{2}\Delta\Gamma t\right) + A_L
      \cos(\Delta{m} t)\right] a^\dag a \right. \\  \left. 
    - \frac{1}{1 + A_L}  \frac{p}{q}
    \left[\sinh\left(\tfrac{1}{2}\Delta\Gamma t\right) - i A_L
      \sin(\Delta{m} t)\right] a^\dag b \right. \\  \left. 
    - \frac{1}{1 - A_L}  \frac{q}{p}
    \left[\sinh\left(\tfrac{1}{2} \Delta\Gamma t\right) + i A_L
      \sin(\Delta{m} t)\right] b^\dag a \right. \\  \left. 
    + \frac{1}{1 - A_L} \left[\cosh\left(\tfrac{1}{2}
        \Delta\Gamma t\right) - A_L \cos(\Delta{m} t)\right] b^\dag
    b\right\}\,.
\end{multline}
The expectation value of $N(t)$ in the state $\ket{n, \bar{n}}$ is
\begin{multline}
  \label{eq:N}
  \left<N(t)\right> = \frac{e^{-\Gamma t}}{1 - A_L^2}
  \left\{\left[\cosh\left(\tfrac{1}{2} \Delta\Gamma t\right) - A_L^2
      \cos(\Delta{m} t)\right] (n + \bar{n}) \right. \\
  \left.  - A_L \left[\cosh\left(\tfrac{1}{2} \Delta\Gamma t\right) -
      \cos(\Delta{m} t)\right] (n - \bar{n})\right\}
\end{multline}
and is depicted in Fig.~\ref{fig:N}.
If we take $A_L = 0$ in Eq.~\eqref{eq:N}, we obtain the case with
preserved $CP$ symmetry, for which we recover the usual result for
mean number of particles
\begin{equation}
  \left<N(t)\right>_{\mathrm{CP}} = 
  \tfrac{1}{2} \left(e^{-\Gamma_S t} + e^{-\Gamma_L t}\right)
  (n + \bar{n})\,.
\end{equation}
The leading part of the difference between $CP$-violated and
$CP$-preserved values is
\begin{equation}
  \left<N(t)\right> - \left<N(t)\right>_{\mathrm{CP}} = -A_L
  \left[\tfrac{1}{2} \left(e^{-\Gamma_S t} + e^{-\Gamma_L t}\right)
    - e^{-\Gamma t} \cos(\Delta{m} t)\right] (n - \bar{n}) +
  O(A_L^2)\,,
\end{equation}
and is shown in Fig.~\ref{fig:N}, too.

\begin{figure}
  \centering
  \includegraphics[width=\columnwidth]{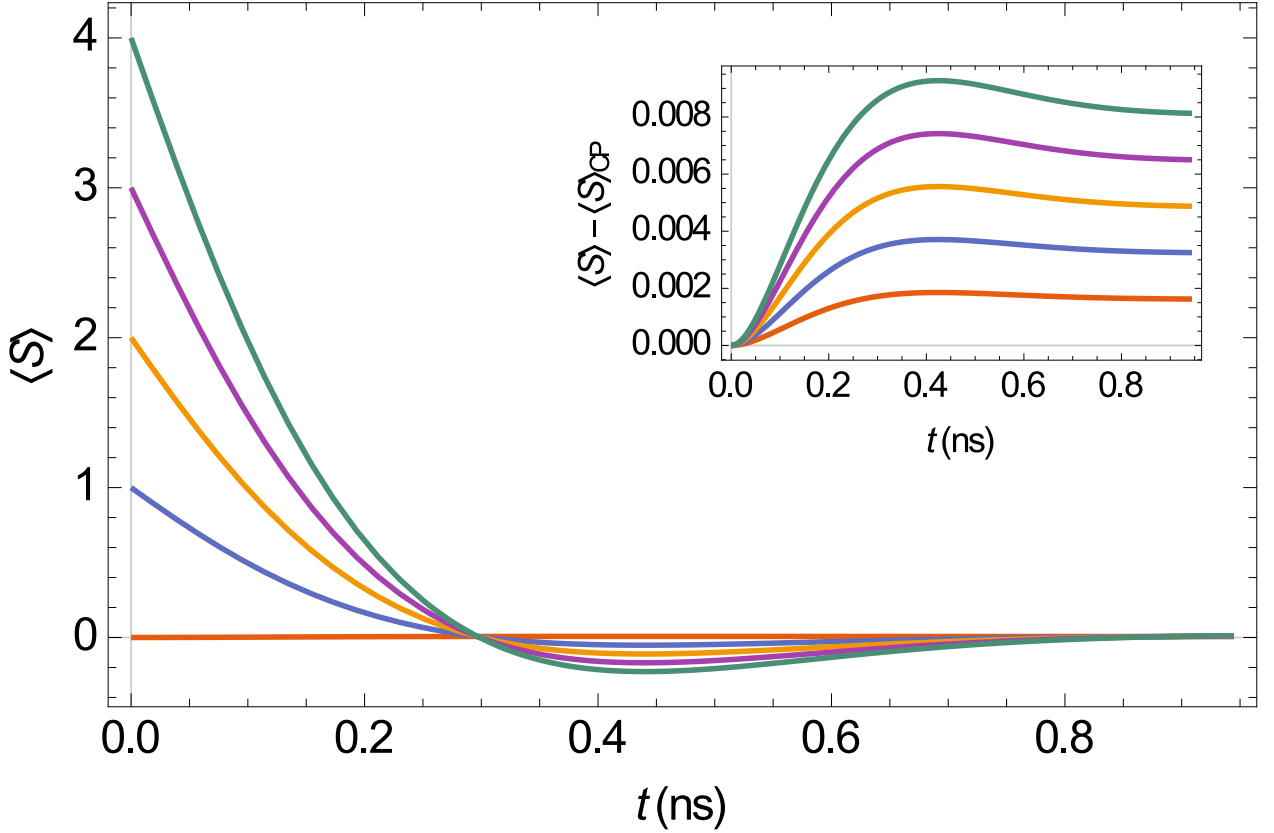}
  \caption{Mean strangeness vs time for $n + \bar{n} = 4$ and $n -
    \bar{n} = 0,\ldots, 4$ (bottom-up) with the strangeness
    oscillation phenomenon explicitly visible when $n - \bar{n} \neq
    0$ and the difference between $\left<S(t)\right>$ for
    $CP$-violated and $CP$-preserved cases vs time for $n + \bar{n} =
    1,\ldots, 5$ and $n - \bar{n} = 0$ (bottom-up).}
  \label{fig:S}
\end{figure}

For strangeness $S$ we have $S(0) = a^\dag a - b^\dag b$, so
$\omega_{aa}(0) = -\omega_{bb}(0) = 1$, $\omega_{ab}(0) =
\omega_{ba}(0) = 0$, and
\begin{multline}
  S(t) = e^{-\Gamma t} \left\{\frac{1}{1 + A_L}\left[A_L
      \cosh\left(\tfrac{1}{2}\Delta\Gamma t\right) + \cos(\Delta{m}
      t)\right] a^\dag a \right.\\
  \left. - \frac{1}{1 + A_L}  \frac{p}{q} \left[A_L
      \sinh\left(\tfrac{1}{2}\Delta\Gamma t\right) - i \sin(\Delta{m}
      t)\right] a^\dag b \right. \\
  \left. - \frac{1}{1 - A_L}  \frac{q}{p} \left[A_L
      \sinh\left(\tfrac{1}{2} \Delta\Gamma t\right) + i \sin(\Delta{m}
      t)\right] b^\dag a \right. \\
  \left.+ \frac{1}{1 - A_L} \left[A_L \cosh\left(\tfrac{1}{2}
        \Delta\Gamma t\right) - \cos(\Delta{m} t)\right] b^\dag
    b\right\}\,.
\end{multline}
The expectation value of $S(t)$ in the state $\ket{n,\bar{n}}$ is therefore
\begin{multline}
  \left<S(t)\right> = \frac{e^{-\Gamma t}}{1 - A_L^2}
  \left\{\left[\cos(\Delta{m} t) - A_L^2 \cosh\left(\tfrac{1}{2}
        \Delta\Gamma t\right)\right] (n - \bar{n}) 
  \right. \\  \left.  
    + A_L \left[\cosh\left(\tfrac{1}{2} \Delta\Gamma t\right) -
      \cos(\Delta{m} t)\right] (n + \bar{n})\right\}
\end{multline}
and is shown in Fig.~\ref{fig:S}.
As previously, when $A_L = 0$ we get the $CP$-preserved value
\begin{equation}
  \left<S(t)\right>_{\mathrm{CP}} = e^{-\Gamma t} \cos(\Delta{m} t) (n
  - \bar{n})\,,
\end{equation}
so the leading term of the difference between $CP$-violated and
$CP$-preserved values is
\begin{equation}
  \left<S(t)\right> - \left<S(t)\right>_{\mathrm{CP}} 
  = A_L \left[\tfrac{1}{2} \left(e^{-\Gamma_S t} + e^{-\Gamma_L t}\right) 
    - e^{-\Gamma t} \cos(\Delta{m} t)\right] (n + \bar{n}) + O(A_L^2)
\end{equation}
and is shown also in Fig.~\ref{fig:S}.

\subsection{Number of $\PKzero$ and $\APKzero$}
\label{sec:K0}

\begin{figure}
  \centering
  \includegraphics[width=\columnwidth]{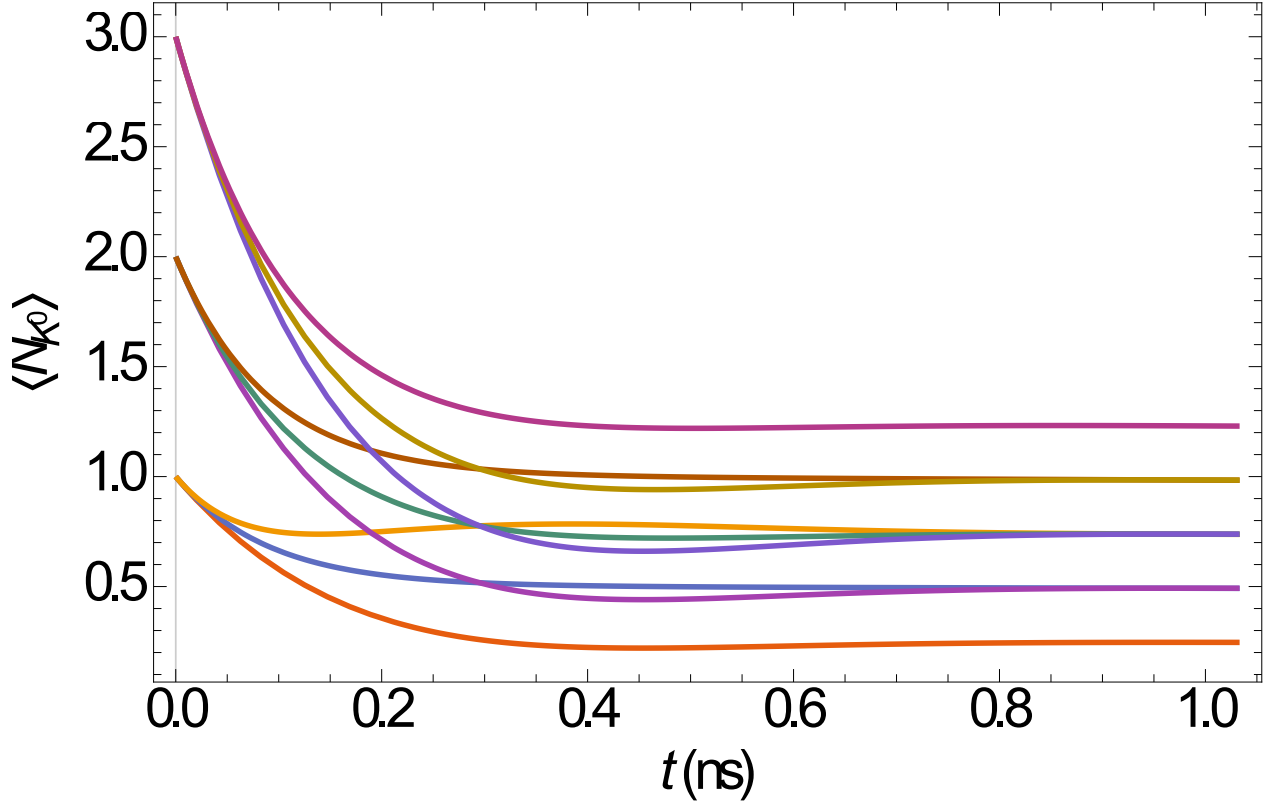}
  \caption{Mean number of $\PKzero$ vs time for $n = 1, 2, 3$ (bundles
    bottom-up) and $\bar{n} = 0, 1, 2$ (bottom-up inside bundles).}
  \label{fig:NPK}
\end{figure}

\begin{figure}
  \centering
  \includegraphics[width=\columnwidth]{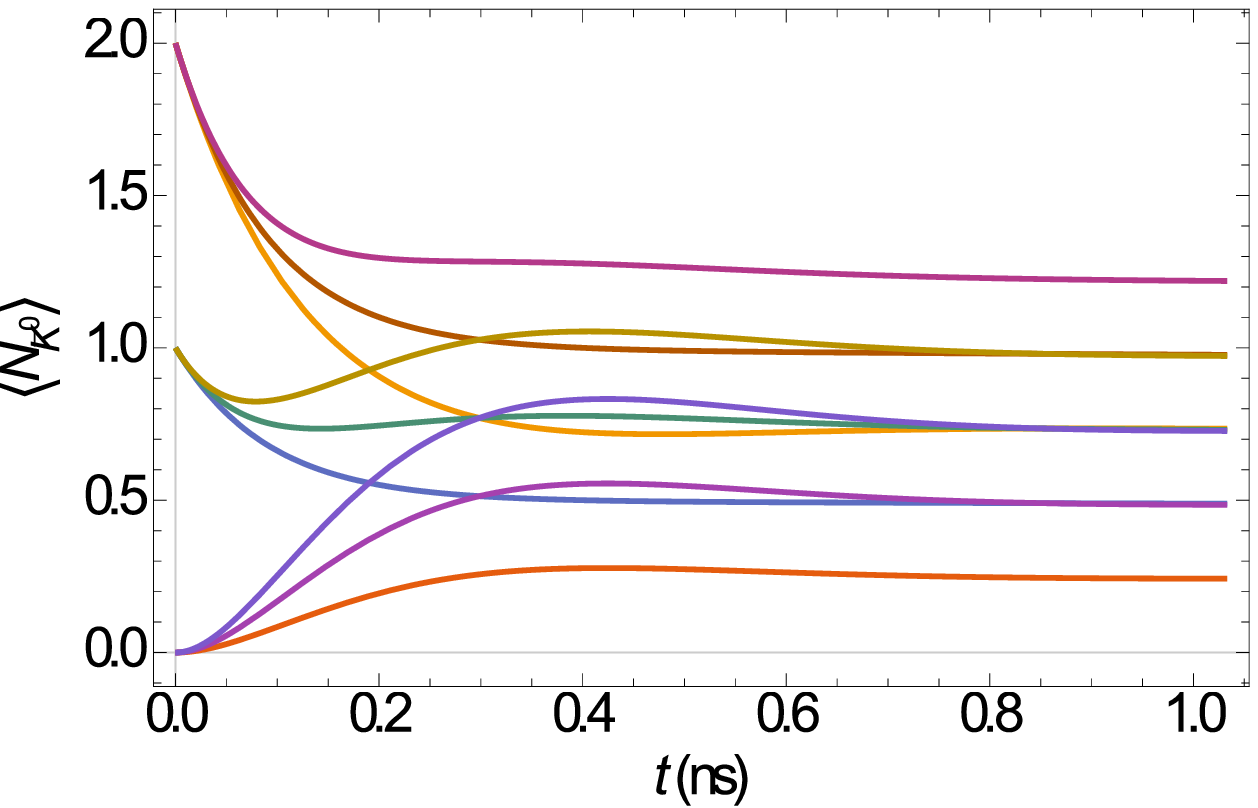}
  \caption{Mean number of $\APKzero$ vs time for $n = 1, 2, 3$
    (bottom-up inside bundles) and $\bar{n} = 0, 1, 2$ (bundles
    bottom-up).}
  \label{fig:NAPK}
\end{figure}

For the numbers of $\PKzero$ and $\APKzero$ we have $N_{\PKzero}(0) =
a^\dag a$ and $N_{\APKzero}(0) = b^\dag b$, so
\begin{subequations}
\begin{multline}
  N_{\PKzero}(t) = \frac{e^{-\Gamma t}}{2} \left\{\vphantom{\frac{1 +
        A_L}{1 - A_L}} \left[\cosh\left(\tfrac{1}{2}\Delta\Gamma
        t\right) + \cos(\Delta{m} t)\right] a^\dag a \right.\\
  \left.- \frac{p}{q} \left[\sinh\left(\tfrac{1}{2}\Delta\Gamma
        t\right) - i \sin(\Delta{m} t)\right] a^\dag b \right.\\
  \left. - \frac{1 + A_L}{1 - A_L}  \frac{q}{p}
    \left[\sinh\left(\tfrac{1}{2} \Delta\Gamma t\right) + i
      \sin(\Delta{m} t)\right] b^\dag a \right.\\
  \left.+ \frac{1 + A_L}{1 - A_L} \left[\cosh\left(\tfrac{1}{2}
        \Delta\Gamma t\right) - \cos(\Delta{m} t)\right] b^\dag
    b\right\}
\end{multline}
and
\begin{multline}
  N_{\APKzero}(t) = \frac{e^{-\Gamma t}}{2} \left\{\frac{1 - A_L}{1 +
      A_L} \left[\cosh\left(\tfrac{1}{2}\Delta\Gamma t\right) -
      \cos(\Delta{m} t)\right] a^\dag a \right.\\
  \left.- \frac{1 - A_L}{1 + A_L}  \frac{p}{q}
    \left[\sinh\left(\tfrac{1}{2}\Delta\Gamma t\right) + i
      \sin(\Delta{m} t)\right] a^\dag b \right.\\
  \left. - \frac{q}{p} \left[\sinh\left(\tfrac{1}{2} \Delta\Gamma
        t\right) - i \sin(\Delta{m} t)\right] b^\dag a \right.\\
  \left.\vphantom{\frac{1 - A_L}{1 + A_L}}+
    \left[\cosh\left(\tfrac{1}{2} \Delta\Gamma t\right) +
      \cos(\Delta{m} t)\right] b^\dag b\right\}\,.
\end{multline}
\end{subequations}
Their mean values are
\begin{subequations}
\begin{equation}
  \langle N_{\PKzero}(t)\rangle = \frac{e^{-\Gamma t}}{2}
  \left\{\vphantom{\frac{1 - A_L}{1 +
        A_L}}\left[\cosh\left(\tfrac{1}{2} \Delta\Gamma t\right) +
      \cos(\Delta{m} t)\right] n 
    + \frac{1 + A_L}{1 - A_L} \left[\cosh\left(\tfrac{1}{2}
        \Delta\Gamma t\right) - \cos(\Delta{m} t)\right]
    \bar{n}\right\}
\end{equation}
and
\begin{equation}
  \langle N_{\APKzero}(t)\rangle = \frac{e^{-\Gamma t}}{2}
  \left\{\frac{1 - A_L}{1 + A_L} \left[\cosh\left(\tfrac{1}{2}
        \Delta\Gamma t\right) - \cos(\Delta{m} t)\right] n 
    \vphantom{\frac{1 - A_L}{1 + A_L}}+
    \left[\cosh\left(\tfrac{1}{2} \Delta\Gamma t\right) +
      \cos(\Delta{m} t)\right] \bar{n}\right\}\,,
\end{equation}
\end{subequations}
respectively.  We show these time evolution in the Figs.~\ref{fig:NPK}
and~\ref{fig:NAPK}.
Notice that after a suitable period of time (approx.\
$8\,\mathrm{ns}$) different initial states can give the same mean
number of $\PKzero$ or $\APKzero$.

If we put $A_L = 0$, we get again $CP$-preserved values
\begin{subequations}
\begin{equation}
  \langle N_{\PKzero}(t)\rangle_{\mathrm{CP}} = 
  \frac{e^{-\Gamma_S t} + e^{-\Gamma_L t}}{2}  \frac{n +
    \bar{n}}{2} + e^{-\Gamma t} \cos(\Delta{m} t) \frac{n -
    \bar{n}}{2}
\end{equation}
and
\begin{equation}
  \langle N_{\APKzero}(t)\rangle_{\mathrm{CP}} = 
  \frac{e^{-\Gamma_S t} + e^{-\Gamma_L t}}{2}  \frac{n +
    \bar{n}}{2} - e^{-\Gamma t} \cos(\Delta{m} t) \frac{n -
    \bar{n}}{2}\,.
\end{equation}
\end{subequations}
The leading term of the differences for the $CP$-violated and
$CP$-preserved vaues are
\begin{subequations}
  \begin{align}
    \langle N_{\PKzero}(t)\rangle - \langle
    N_{\PKzero}(t)\rangle_{\mathrm{CP}} &= A_L \left[\tfrac{1}{2}
      \left(e^{-\Gamma_S t} + e^{-\Gamma_L t}\right) 
      - e^{-\Gamma t} \cos(\Delta{m} t)\right] \bar{n} + O(A_L^2)\,,\\
    \langle N_{\APKzero}(t)\rangle - \langle
    N_{\APKzero}(t)\rangle_{\mathrm{CP}} &= -A_L \left[\tfrac{1}{2}
      \left(e^{-\Gamma_S t} + e^{-\Gamma_L t}\right) 
      - e^{-\Gamma t} \cos(\Delta{m} t)\right] n + O(A_L^2)\,.
  \end{align}
\end{subequations}

\subsection{Number of $\PKshort$ and $\PKlong$}
\label{sec:KsKl}

Let us return to the short and long living states of neutral kaon,
$\ket{\PKshort}$ and $\ket{\PKlong}$, respectively.  If we define
another two pairs of annihilation and creation operators, such that
\begin{subequations}
    \begin{align}
    \ket{\PKshort} &= c_S^\dag \ket{0}\,, \\
    \ket{\PKlong} &= c_L^\dag \ket{0}\,,
  \end{align}
\end{subequations}
we can find, with use of Eq.~\eqref{eq:KSL}, that
\begin{subequations}
  \begin{align}
    c_S &= p^* a + q^* b\,,\\
    c_L &= p^* a - q^* b\,.
  \end{align}
\end{subequations}
The operators $c_S$ and $c_L$ fulfill almost the usual canonical
commutation relations with the following exception:
\begin{equation}
  \label{eq:SLccr}
  [c_S, c_L^\dag] = [c_L, c_S^\dag] = A_L\,.
\end{equation}
This reflects the fact that $\ket{\PKshort}$ and $\ket{\PKlong}$ are
not orthogonal, see Eq.~\eqref{eq:braketSL}.

Now we define multiparticle states for $\PKshort$ and $\PKlong$ as
\begin{subequations}
  \begin{align}
    \ket{\#\PKshort = n} \equiv \ket{n_S} &=
    \frac{(c_S^\dag)^{n}}{\sqrt{n!}} \ket{0}\,,\\
    \ket{\#\PKlong = n} \equiv \ket{n_L} &=
    \frac{(c_L^\dag)^{n}}{\sqrt{n!}} \ket{0}\,.
  \end{align}
\end{subequations}
These states can be expressed in terms of the sates $\ket{n, k}$
as follows
\begin{subequations}
  \begin{align}
    \label{eq:nSnn}
    \ket{n_S} &= \sum_{k=0}^{n} \sqrt{\binom{n}{k}} p^{n-k} q^k 
    \ket{n-k,k}\,, \\
    \label{eq:nLnn}
    \ket{n_L} &= \sum_{k=0}^{n} (-1)^k \sqrt{\binom{n}{k}} p^{n-k} q^k 
    \ket{n-k,k}\,
  \end{align}
\end{subequations}

\begin{widetext}
For the number of $\PKshort$, $N_{\PKshort}(0) = c_S^\dag c_S = \frac{1 +
  A_L}{2} a^\dag a + \frac{1 - A_L}{2}  \frac{p}{q} a^\dag b +
\frac{1 + A_L}{2}  \frac{q}{p} b^\dag a + \frac{1 - A_L}{2}
b^\dag b$, so
\begin{multline}
  N_{\PKshort}(t) = \frac{e^{-\Gamma t}}{2} \left\{\frac{1}{1 + A_L}
    \left[e^{-\tfrac{1}{2}\Delta\Gamma t} + A_L^2
      e^{\tfrac{1}{2}\Delta\Gamma t} + 2 A_L \cos(\Delta{m} t)\right]
    a^\dag a \right. \\  \left. 
    + \frac{1}{1 + A_L}  \frac{p}{q}
    \left[e^{-\tfrac{1}{2}\Delta\Gamma t} - A_L^2
      e^{\tfrac{1}{2}\Delta\Gamma t} + 2 i A_L \sin(\Delta{m}
      t)\right] a^\dag b \right.\\  \left.
    + \frac{1}{1 - A_L} 
    \frac{q}{p}\left[e^{-\tfrac{1}{2}\Delta\Gamma t} - A_L^2
      e^{\tfrac{1}{2}\Delta\Gamma t} - 2 i A_L \sin(\Delta{m}
      t)\right] b^\dag a \right.\\  \left. 
    + \frac{1}{1 - A_L}\left[e^{-\tfrac{1}{2}\Delta\Gamma t} +
      A_L^2 e^{\tfrac{1}{2}\Delta\Gamma t} - 2 A_L \cos(\Delta{m}
      t)\right] b^\dag b\right\}\,.
\end{multline}
Similarly, for the number of $\PKlong$, $N_{\PKlong}(0) = c_L^\dag c_L =
\frac{1 + A_L}{2} a^\dag a - \frac{1 - A_L}{2}  \frac{p}{q}
a^\dag b - \frac{1 + A_L}{2}  \frac{q}{p} b^\dag a + \frac{1 -
  A_L}{2} b^\dag b$, and
\begin{multline}
  N_{\PKlong}(t) = \frac{e^{-\Gamma t}}{2} \left\{\frac{1}{1 + A_L}
    \left[e^{\tfrac{1}{2}\Delta\Gamma t} + A_L^2
      e^{-\tfrac{1}{2}\Delta\Gamma t} + 2 A_L \cos(\Delta{m} t)\right]
    a^\dag a \right. \\  \left. 
    - \frac{1}{1 + A_L}  \frac{p}{q}
    \left[e^{\tfrac{1}{2}\Delta\Gamma t} - A_L^2
      e^{-\tfrac{1}{2}\Delta\Gamma t} - 2 i A_L \sin(\Delta{m}
      t)\right] a^\dag b \right.\\
  \left.- \frac{1}{1 - A_L} 
    \frac{q}{p}\left[e^{\tfrac{1}{2}\Delta\Gamma t} - A_L^2
      e^{-\tfrac{1}{2}\Delta\Gamma t} + 2 i A_L \sin(\Delta{m}
      t)\right] b^\dag a \right.\\  \left.
    + \frac{1}{1 - A_L}\left[e^{\tfrac{1}{2}\Delta\Gamma t} +
      A_L^2 e^{-\tfrac{1}{2}\Delta\Gamma t} - 2 A_L \cos(\Delta{m}
      t)\right] b^\dag b\right\}\,.
\end{multline}
\end{widetext}
Now we are prepared to find the expectation values of
$N_{\PKshort}(t)$ and $N_{\PKlong}(t)$ in the states $\ket{n_S}$ and
$\ket{n_L}$, respectively.

First, observe that
\begin{subequations}
  \begin{align}
    a^\dag b \ket{n_S} &= \frac{q}{p} a^\dag a \ket{n_S}\,,&
    a^\dag b \ket{n_L} &= -\frac{q}{p} a^\dag a \ket{n_L}\,,\\
    b^\dag a \ket{n_S} &= \frac{p}{q} b^\dag b \ket{n_S}\,,&
    b^\dag a \ket{n_L} &= -\frac{p}{q} b^\dag b \ket{n_L}\,.
  \end{align}
\end{subequations}
Thus
\begin{subequations}
  \begin{equation}
    \label{eq:Ns}
    N_{\PKshort}(t) \ket{n_S} = \frac{e^{-\Gamma t}}{1 - A_L^2}
    \left[\left(e^{-\frac{1}{2} \Delta\Gamma t} - A_L^2 e^{i \Delta m
          t}\right) N(0)
      - A_L \left(e^{-\frac{1}{2} \Delta\Gamma t} - e^{i \Delta m
          t}\right) S(0)\right] \ket{n_S}
  \end{equation}
  and
  \begin{equation}
    \label{eq:Nl}
    N_{\PKlong}(t) \ket{n_L} = \frac{e^{-\Gamma t}}{1 - A_L^2}
    \left[\left(e^{\frac{1}{2} \Delta\Gamma t} - A_L^2 e^{-i \Delta m
          t}\right) N(0)
      - A_L \left(e^{\frac{1}{2} \Delta\Gamma t} 
        - e^{-i \Delta m t}\right) S(0)\right] 
    \ket{n_L}\,.
  \end{equation}
\end{subequations}
Of course $\ket{n_S}$ and $\ket{n_L}$ are eigenvectors of $N(0)$:
\begin{subequations}
  \begin{align}
    N(0) \ket{n_S} &= n \ket{n_S}\,,\\
    N(0) \ket{n_L} &= n \ket{n_L}\,,
  \end{align}
\end{subequations}
but they are not eigenvectors of $S(0)$ since
\begin{subequations}
\label{eq:S0nSL}
  \begin{align}
    S(0) \ket{n_S} &= n \ket{n_S} 
    - 2 \sum_{k=0}^n k \sqrt{\binom{n}{k}} p^{n-k} q^k \ket{n-k,k}\,,\\
    S(0) \ket{n_L} &= n \ket{n_L} 
    - 2 \sum_{k=0}^n (-1)^k k \sqrt{\binom{n}{k}} p^{n-k} q^k \ket{n-k,k}\,.
  \end{align}
\end{subequations}

Notice that in view of Eq.~\eqref{eq:S0nSL} the states $\ket{n_S}$ and
$\ket{n_L}$ are \emph{not} eigenvectors of $N_{\PKshort}$ and
$N_{\PKlong}$ respectively, as one can naïvely expect.  However,
the expectation values of $S(0)$ in the states $\ket{n_S}$ and
$\ket{n_L}$ are well established and can be calculated as
\begin{subequations}
  \begin{align}
    \bra{n_S} S(0) \ket{n_S} &= A_L n\,,\\
    \bra{n_L} S(0) \ket{n_L} &= A_L n\,,
  \end{align}
\end{subequations}
because
\begin{equation}
  \sum_{k=0}^n k \binom{n}{k} |p|^{2(n-k)} |q|^{2k} 
  = n |q|^2 \left(|p|^2 + |q|^2\right)^{n-1} = n |q|^2\,.
\end{equation}
In conclusion, we get that the mean number of $\PKshort$ and $\PKlong$
in the states $\ket{n_S}$ and $\ket{n_L}$, respectively, evolves in
time according to the Geiger-Nutall law
\begin{subequations}
  \begin{align}
    \bra{n_S}N_{\PKshort}(t)\ket{n_S} &= n e^{-\Gamma_S t}\,,\\
    \bra{n_L}N_{\PKlong}(t)\ket{n_L} &= n e^{-\Gamma_L t}\,.
  \end{align}
\end{subequations}

The states $\ket{n_S}$ and $\ket{n_L}$ are not orthogonal, thus we
expect nonzero expectation values of $N_{\PKshort}(t)$ and
$N_{\PKlong}$ in the states $\ket{n_L}$ and $\ket{n_S}$, respectively.
Indeed,
\begin{subequations}
  \begin{equation}
    \label{eq:Ns}
    N_{\PKshort}(t) \ket{n_L} = \frac{e^{-\Gamma t} A_L}{1 - A_L^2}
    \left[\left(e^{-i \Delta m t} - A_L^2 e^{\frac{1}{2} \Delta\Gamma
          t}\right) S(0)
      - A_L \left(e^{-i \Delta m t} - e^{\frac{1}{2} \Delta\Gamma
          t}\right) N(0)\right] \ket{n_L}
  \end{equation}
  and
  \begin{equation}
    \label{eq:Nl}
    N_{\PKlong}(t) \ket{n_S} = \frac{e^{-\Gamma t} A_L}{1 - A_L^2} 
    \left[\left(e^{i \Delta m t} - A_L^2 e^{-\frac{1}{2} \Delta\Gamma t}\right) 
      S(0)
      - A_L \left(e^{i \Delta m t} 
        - e^{-\frac{1}{2} \Delta\Gamma t}\right) N(0)\right] 
    \ket{n_S}\,,
  \end{equation}
\end{subequations}
thus,
\begin{subequations}
  \begin{align}
    \bra{n_L}N_{\PKshort}(t)\ket{n_L} &= n A_L^2 e^{-\Gamma_L t}\,,\\
    \bra{n_S}N_{\PKlong}(t)\ket{n_S} &= n A_L^2 e^{-\Gamma_S t}\,.
  \end{align}
\end{subequations}
So, finally we see that we can detect a fraction of order $A_L^2$ of
the other flavor in multiparticle short- or long-lived neutral kaon
states.

\section{Conclusions}
\label{sec:concl}

We have shown that it is possible to formulate the consistent and
probability-preserving description of the $CP$-symmetry-violating
evolution of a system containing any number of decaying particles.
This has been done within the framework of quantum mechanics of open
systems based on the approach developed in
\cite{caban05a,smolinski14}.  To achieve this aim we have considered
master equations built up from creation and annihilation operators
which generate dynamical semigroups that can describe the exponential
decay and flavour oscillations for a system of many particles.  It
should be noted that this dynamical semigroup was used for the
description of the entire system of unstable particles, and not only
for decoherence effects, as it was done in, e.g.,
\cite{bertlmann06,benatti96,*benatti97a,*benatti97b,*benatti98a,
  *benatti98,*benatti01,*benatti98b:full}.

We have used the fact that the decay of the particle can be regarded
as a Markov process (the probability of decay of a particle is
constant, so it does not depend on its history) and we have solved
explicitly the Kossakowski-Lindblad master equation for a system of
particles with violated $CP$ symmetry.  To avoid working with infinite
numbers of density matrix elements we performed calculations in the
Heisenberg picture of quantum mechanics.  This allowed us to deal with
a low-dimensional system of linear differential equations for
evolution of \emph{any} observable bilinear in creation and
annihilation operators.  The choice of a concrete observable is then
reduced to the proper choice of initial conditions for the system of
differential equations.  To show explicitly the effectiveness of the
introduced approach we have calculated the evolution of the
observables which are most interesting from the physical point of
view.  We have also found the evolution of mean values of these
observables as well as their lowest order difference with the
$CP$-preserved values.

It seems to us that the presented analysis of time evolution of
neutral kaons could be applied to the description of EPR states.  Of
course one must construct such a state by labeling annihilation and
creation operators either by a discrete index, as it is done in
quantum optics, or by a continuous parameter, as it is done in quantum
field theory (beware that in the latter case annihilation and creation
operators are actually operator valued distributions).

Moreover, we think that our approach would be interesting and helpful
also in analysis of mechanical systems and electric circuits.  This
follows from the fact that there is a~strong analogy with
nonreversible classical mechanics and quantum systems with $CP$
violation \cite{rosner96,*cocolicchio98,*kostelecky01,*caruso11}.

Finally, we would like to point out that all the presented results are
also valid for neutral $B$ mesons after appropriate change of notation
and values of physical quantities.  This follows from the fact that
$B$ mesons evolve according to the same scheme as kaons.

\begin{acknowledgments}
  This work is supported by the Polish National Science Centre under
  Contract No.\ 2014/15/B/ST2/00117 and by the University of Lodz.
\end{acknowledgments}

%
\end{document}